\documentstyle[12pt]{article}
\renewcommand{\thefootnote}{\fnsymbol{footnote}}
\newcommand{\EQ}{\begin{equation}}
\newcommand{\EN}{\end{equation}}
\newcommand{\bea}{\begin{eqnarray}}
\newcommand{\ena}{\end{eqnarray}}

\newcommand{\uda}{\nearrow \kern-1em \searrow}
\newcommand{\half}{{1 \over2}}

\begin{document}

\topmargin 0pt
\oddsidemargin 5mm

\begin{titlepage}
\setcounter{page}{0}
\begin{flushright}
OU-HET 202 \\
LMU-TPW 94-13 \\
October, 1994
\end{flushright}

\begin{center}
{\Large On Bogoliubov Transformation of Scalar Wave
  Functions \\ in De Sitter Space}

\vspace{2cm}
{\large Haru-Tada Sato
\footnote{Fellow of the Japan Society for the
    Promotion of Science, On leave of absence from Osaka University \\
    \phantom{mil} e-mail address: hsato@phys.wani.osaka-u.ac.jp}}\\
{\em Sektion Physik der Universit{\"a}t M{\"u}nchen, \\
Theresienstrasse 37, D-80333 M{\"u}nchen, Germany} \\
{\large Hisao Suzuki
\footnote{e-mail address: suzuki@phys.wani.osaka-u.ac.jp}}\\
{\em Department of Physics,
Osaka University \\ Toyonaka, Osaka 560, Japan} \\
\end{center}

\centerline{{\bf{Abstract}}}
We discuss the Bogoliubov transformation of the scalar wave functions caused
by the change of coordinates in 4 dimensional de Sitter space. It is shown
that the exact Bogoliubov coefficients can be obtained from the global
coordinates to the static coordinates where there exist manifest horizon.
We consider two type of global coordinates. In one global coordinates, it
is shown that the Bogoliubov transformation to the static coordinates can be
expressed by the discontinuous integral of Weber and Schafheitlin.
The positive and negative energy states in the global coordinates degenerate
in the static coordinates. In the other global coordinates, we obtain the
Bogoliubov coefficients by using the analytic continuation of the
hypergeometric functions in two variables.

We also discuss the relation between two type of global coordinates and
find an integral relation between the mode functions.
\end{titlepage}
\newpage
\renewcommand{\thefootnote}{\arabic{footnote}}
\setcounter{footnote}{0}
\section{Introduction}
Quantum field theories in curved space time have been investigated
extensively. One of the celebrating result is the existence of Hawking
radiations\cite{Haw}. Although true nature of the radiation should be
revealed when we include the effect of the back reaction, some of the
difficulty of the treatment arise from the fact that the quantum effect
such as in Schwarzschild metric cannot be expressed by well-known special
functions where analytic properties can be extracted. A simple example of
the metric which possess  horizons arises in the cosmological situation such
as de Sitter space. In this metric, it was shown that the wave functions of
the scalar fields can be solved by special functions in several
coordinates\cite{Nac}\cite{CT}\cite{Tag}\cite{RP}. Hawking radiation has
been analyzed by several arguments. One way is the use of Kruskal
coordinates and the analytic continuation of the Green
functions.~\cite{GH}. Later, an explicit scattering state has been
considered  using Unruh`s observation\cite{Unr} and Gibbons-Hawking
vacuum.\cite{RP} The other interesting method is to consider the Bogoliubov
transformation of the mode functions between the static coordinates and
the global coordinates where there is no loss of information.\cite{Lap}
Namely, we can define the vacuum in the static coordinates by use of the
vacuum of the global coordinates. this program has been discussed by using
high frequency approximation and find the factor showing the existence
of the radiation. In principle, the de Sitter space seems to be one of the
rare example that we can obtain the exact Bogoliubov coefficients
analytically because the modes functions in both coordinates can be expressed
by special functions. The coefficients should contain the information of
the effects of various parameter such as masses to Hawking radiation
without using S-wave approximation.

Our aim of this paper is to evaluate the Bogoliubov coefficients of the
transformation exactly. We treat two type of globally defined coordinates
where there is no horizon. We obtain the transformation functions of the
scalar wave functions to the static coordinates. The Bogoliubov
transformation from one of the global coordinates can be expressed by the
so-called discontinuous integral of Weber and Schafheitlin type\cite{HTFI}.
Using this formula we show that the we cannot define the well defined vacua
in the static coordinates. Actually,the non-analyticity of mode functions
is known for lower dimensional examples such as Rindler space. Our case
must be a four-dimensional example of such known phenomena. As for the
Bogoliubov coefficient from the other global coordinates, we have to use
the analytic continuation of the mode functions inside the horizon. We
evaluate the coefficients in terms of the hypergeometric functions in
two variables. For this purpose, we prove some formula for these functions.

We also discuss the relation between the mode functions between two type of
the global coordinates and find an integral formula. We comment on the
relation of our results and the previous calculation. It seems interesting
that these transformations can be expressed by various integral formula
including bi-linear of the special functions.
\section{Notation}

The coordinates of  de Sitter space are obtained by various
parameterization of the flat space-time of the coordinates $z_a$
$a=0,2,\ldots,4$ satisfying
\bea
-z_0^2 + z_1^2 + z_2^2+z_3^2+z_4^2 = 1,
\ena
where we have normalized the radius to one.
A global parameterization of this coordinates is given by
\bea
z_0 &=& \sinh t_1 + {1 \over 2}e^{t_1} r_1^2, \nonumber\\
z_4 &=& \cosh t_1 - {1\over 2} e^{t_1} r_1^2, \nonumber\\
z_1 &=& e^{t_1} r_1\sin\theta \sin\phi, \nonumber\\
z_2 &=& e^{t_1} r_1\sin\theta \cos\phi, \nonumber\\
z_3 &=& e^{t_1} r_1\cos\theta.
\ena
The conformal time is given by
\bea
\eta_1 = - e^{-t_1}, \qquad - \infty < \eta_1 < 0,
\ena
by which the metric is expressed as
\bea
(ds)^2 = {1 \over \eta_1^2}[ d\eta_1^2 - dr_1^2 - r_1^2(d\theta^2 +
\sin\theta^2 d\phi^2)].
\ena
We shall call this coordinates as type-I global coordinates. In this
coordinates, the Klein-Gordon equation for a scalar field
$g^{\mu\nu}\nabla_\mu\nabla_\nu \phi + m^2 \phi =0$ can be solved in terms
of the Bessel functions as\cite{Nac}
{\bea
{\phi_{k,l,m}^{I,\pm\nu}(\eta_1, r_1, \theta, \phi) = (-k\eta_1)^{3/2}
 J_{\pm\nu}(-k\eta_1 )(kr_1 )^{-1/2} J_{l+1/2}(kr_1)Y_{l,m}(\theta,\phi )},
\ena}
where we have chosen $J_{\pm\nu}$ to be two independent solutions for
the time $\eta_1$. $Y_{l,m}(\theta,\phi )$ is the spherical harmonics and
$\nu = (9/4 -m^2)^{1/2}$ .

Another global coordinates is given by
\bea
z_0 &=& \sinh t_2 , \nonumber\\
z_4 &=& \cosh t_2 \cos\chi, \nonumber\\
z_1 &=& \cosh t_2 \sin\chi \sin\theta\sin\phi, \nonumber\\
z_2 &=& \cosh t_2 \sin\chi \sin\theta\cos\phi, \nonumber\\
z_3 &=& \cosh t_2\sin\chi\cos\theta,
\ena
which we denote by type-II global coordinates.
We introduce the conformal time in this system as
\bea
\tan{\eta_2 \over 2} = e^{t_2}, \qquad 0 \leq \eta_2 <\pi,
\ena
so that the  metric is expressed as
\bea
ds^2 = \sin^{-2}\eta_2 [d\eta_2^2 - d\chi^2 - \sin^2\chi(d\theta^2+
\sin^2\theta d\phi^2)].
\ena
In this coordinates, the mode function are expressed as\cite{CT}\cite{Tag}
\bea
\phi_{N,l,m}^{II,\pm\nu}(\eta_2,\chi,\theta,\phi) = (\sin\eta_2)^{3/2}
 P^{\pm\nu}_{N+1/2}(\cos\eta_2)C_{N-l}^{l+1}(\cos\chi)(\sin\chi)^l
Y_{l,m}(\theta,\phi),
\ena
where $P^\mu_\nu(z)$ is the associated Legendre function and $C_n^\nu(z)$
denotes Gegenbauer`s polynomial. We can also choose
$P^{\nu}_{N+\half}(\cos\eta_2)$ and $Q^{\nu}_{N+\half}(\cos\eta_2)$ as
two independent solutions for the mode functions with respect to the
variable $\eta_2$.

As the third coordinates, we consider the static coordinates defined by
\bea
z_0 &=& (1-r^2)^{1/2}\sinh t , \nonumber\\
z_4 &=& (1-r^2)^{1/2} \cosh t, \nonumber\\
z_1 &=& r \sin\theta\sin\phi, \nonumber\\
z_2 &=& r \sin\theta\cos\phi, \nonumber\\
z_3 &=& r \cos\theta.
\ena
This coordinates covers the half of the de Sitter space with $z_0 + z_4
> 0$. In this coordinates, the metric can be written as
\bea
ds^2 = (1-r^2)dt^2 - (1-r^2)^{-1}dr^2 - r^2(d\theta^2 + \sin\theta^2d \phi^2).
\ena
The line element possesses a coordinates singularity at $r=1$, which is
the event horizon for an observer at $r=0$. The solution of the scalar
wave functions are written by the hypergeometric functions\cite{RP};
\bea
 \phi_{w,l,m}^{S,(\pm)}(t,r,\theta,\phi)
  = e^{\pm i\omega t}r^l(1-r^2)^{i\omega/2} F(a,b;c;r^2),
\ena
where
\bea
a,b &=& \half c + {i\omega \over2} \pm {\nu \over 2},
\nonumber\\
c &=& l + {3 \over 2},
\ena
and $F(a,b;c;z)$ is the hypergeometric function.  We have omitted the
radial functions which diverges at $r=0$ as in $(2.9)$.
\setcounter{equation}{0}
\section{Bogoliubov transformation from the type-I coordinates to
the static coordinates}

We are going to argue the relations among these coordinates. To begin
with, we consider the transformation between the static coordinates and
the type-I global coordinates.  By comparing $(2.2)$ and $(2.10)$, we can
find the relation of the coordinates $\eta_1,r_1$ and $t, r$ as
\bea
             r &=& - {r_1 \over \eta_1}, \nonumber\\
             t &=& {1 \over 2}\ln(\eta_1^2 - r^2_1),
\ena
We take the Bogoliubov transformation from $(2.5)$ to $(2.12)$ to be of
the form;
\bea
\phi^{S(\pm)}_{\omega,l,m}(t,r) =
\int_0^\infty a^{(\pm)}_{\omega}(k,\nu)\phi^{I,\nu}_{k,l,m}(\eta_1,r_1)dk,
\ena
where $a(k)$ is a function to be determined. The coefficient $a^{\pm}$ as
a function of $k$ can be obtained by considering the Killing vector
${\partial \over \partial t}$. Note that the functions
$\phi^{S(\pm)}$ are the eigenfunction of this killing vector which can
be written in the global coordinates as
\bea
{\partial \over \partial t} = \eta_1 {\partial \over \partial \eta_1}
  + r {\partial \over \partial r}.
\ena
Since the mode function $(2.5)$ is the function of $k\eta_1$ and $kr_1$,
the action of the killing vector to $(3.2)$ is given by
$k{ \partial \over \partial k}$ in the integral. By using the partial
integration, we get the following equation for $a(k)$;
\bea
        - {d \over dk}(k a^\pm(k)) = \pm iw a^\pm(k).
\ena
The solution of this equation is simply given by
\bea
            a^{(\pm)}(k) = c k^{\mp iw-1},
\ena
where $c$ is a coefficient independent of $k$. The integral of the form
$(3.2)$ with $(3.5)$ is known as the discontinuous integral
of Weber and Schafheitlin\cite{HTFI}, which is
\bea
&{}&2^\rho a^{-\mu} b^{\mu-\rho+1}
   {\Gamma(\mu+1)\Gamma(\half+\half\nu+\half\rho -\half\mu) \over
    \Gamma(\half+\half\nu+\half\mu-\half\rho)}
   \int_0^\infty J_\mu(at)J_\nu(bt)t^{-\rho} \nonumber\\
&{}& \quad  F(\half + \half\nu + \half\mu-\half\rho,
     \half+\half\mu-\half\nu-\half\rho;\nu+1;{a^2 \over b^2}).
\ena
By comparing this integral formula to $(3.2)$, we have
\bea
 \phi^{S,-} = 2^{-i\omega } {\Gamma(l+{3\over2})
   \Gamma(\half(1\pm\nu-i\omega -l-\half)) \over \Gamma(\half(\pm\nu
   +l +\half + i\omega))} \int_0^\infty dk k^{iw-1} \phi^{I,\pm\nu}.
\ena
Note that the two independent solutions in global coordinates are transformed
to an identical mode function in the static coordinates. Namely, the positive
and negative energy wave functions are degenerate in the static coordinates.
Correspondingly, another mode function $\phi^+$ in the static coordinates
can not be represented by the Bogoliubov transformation from the type-I
global coordinates. This mode can be obtained by the analytic continuation
from inside the horizon $r^2 > 1$.
%
\setcounter{equation}{0}
\section{The type-II to the static coordinates}

We next consider the Bogoliubov transformation the type-II coordinates
and the static coordinates, which is discussed in ref.\cite{Lap} using
approximations. We consider the Bogoliubov transformation of the form
\bea
\phi^{S,\pm}(t,r,\theta,\varphi) = \sum_{N\geq l} [ A_N^\pm
P^\nu_{N+\half}(\cos\eta_2) + B_N^\pm
Q_{N+\half}^\nu(\cos\eta_2)]C^{l+1}_{N-l}(\cos\chi)(\sin\chi)^l
Y_{l,m}(\theta, \varphi),
\ena
The change of variables is given by
\bea
t&=& {1 \over2}[ \ln(x-y) - \ln(x+y)], \nonumber\\
r&=& \bigl( {1 - x^2 \over 1 - y^2} \bigr)^{1/2},
\ena
where $x = \cos\chi$, $y= \cos\eta_2$. As in the previous consideration,
we can construct a recursion relation of the coefficients $A^\pm_N$ and
$B^\pm_N$ by considering the killing vector
\bea
{\partial \over \partial t} = - (1-x^2)y { \partial \over \partial x}
- (1-y^2)x{\partial \over \partial y}.
\ena
Expressing the Gegenbauer polynomial in terms of the Legendre functions
and using recursion relations of the Legendre functions both for variables
$x$ and $y$, we obtain the following difference equations;
\bea
{(N+\nu+{3\over2})(N+l+2) \over 2(N+2)}C_{N+1}^\pm &=&\mp i \omega
C_N^\pm + {(N-\nu+\half)(N-l) \over 2N}C_{N-1}^\pm ,\\
C^\pm_{l+1} &=& \mp i\omega {l+2 \over (l+1)(l+\nu+{3\over2})}C^\pm_l,
\ena
where $C_N^\pm$ denotes both $A_N^\pm$ and $B_N^\pm$. Hence, after obtaining
one of the coefficients $A_l^\pm$ and $B_l^\pm$, we can determine the other
coefficients through these recursion formulae $(4.4)$ and (4.5). Note also
that the ratio of $A_N$ and $B_N$ does not depend on $N$ because both
coefficients obey an identical recursion relation. We can determine the
coefficient for $N=l$ by using the orthogonality of Gegenbauer's polynomial
and find the mode function
$G^{\pm}_l(\eta_2)\equiv A^\pm_lP^\nu_{l+\half}(\cos\eta_2) +
B^\pm_l Q_{l+\half}^\nu(\cos\eta_2)$  can be expressed by
\bea
G^\pm_l(\eta_2)
&=& {\Gamma(l+2) \over \pi^\half \Gamma(l+{3\over2})} \int_0^\pi
    \phi^{S,\pm}(t,r)(\sin\chi)^{l+2} \nonumber\\
&=& {\Gamma(l+2) \over \pi^\half \Gamma(l+{3\over2})}(\sin\eta_2)^{-l-iw}
    \int_0^\pi d\chi
(\cos\eta_2\mp\cos\chi)^{i\omega}(\sin\chi)^{2l+2}F(a,b;c;{\sin^2\chi
  \over \sin^2\eta_2}) \nonumber\\
&=& {\Gamma(l+2) \over \pi^\half \Gamma(l+{3\over2})}(\sin\eta_2)^{-l-iw}
 \int_0^{\pi \over 2} d\chi
[(\cos\eta_2+\cos\chi)^{i\omega}+ (\cos\eta_2 - \cos\chi)^{i\omega}]
\nonumber\\
&{}& \times (\sin\chi)^{2l+2}F(a,b;c;{\sin^2\chi
  \over \sin^2\eta_2}),
\ena

In order to find explicit Bogoliubov coefficients, we have to use
analytic continuation because the integral over $\chi$ does not preserve
the condition $r^2<1$. The integral formula of this type can not be found
in the literature so we prove some formula. For this purpose, making use of
the following representation\cite{HTFII};
\bea
(1+z^\half)^{-2a} + (1-z^\half)^{-2a} = 2 F(a,a+\half;\half;z),
\ena
We can express the RHS of $(4.6)$ in the following form;
\bea
G_l(\eta_2) &=& {2 \Gamma(l+2) \over \pi^\half \Gamma(l+{3\over 2})}
(\sin\eta_2)^{-l-iw}(\cos\eta_2)^{i\omega} \nonumber\\
&{}& \times \int_0^{\pi\over2} d\chi
(\sin\chi)^{2l+2} F(-
{i\omega \over 2}, -{i \omega \over 2} + \half; \half; {\cos^2\chi
  \over \cos^2\eta_2})F(a,b;c;{\sin^2\chi \over \sin^2 \eta_2})
\nonumber\\
&=&{ \Gamma(l+2) \over \pi^\half \Gamma(l+{3\over 2})}
(1-x^2)^{\half{(-l-iw)}}x^{i\omega} \nonumber\\
&{}&\times
\int_0^1 dt
t^{-\half}(1-t)^{c-1} F(-
{i\omega \over 2}, -{i \omega \over 2} + \half; \half; {t
  \over x^2})F(a,b;c;{1-t \over 1-x^2}),
\ena
where we have set $\cos^2\chi =t$ and  $\cos\eta_2 = x$. Note that the
integral of the form can be expressed by the hypergeometric functions in
two variables \cite{HTFFIIIINT} as
\bea
G_l = x^{iw}(1-x^2)^{-\half(i\omega+l)} F_3(-{i\omega \over 2},a, -{i\omega
  \over 2} +\half,b;c+\half;{1 \over x^2},{1 \over 1-x^2}),
\ena
where the function
$F_3(\alpha,\alpha^\prime,\beta,\beta^\prime,\gamma,x,y)$ is defined
by\cite{HTFFIIIDEF}
\bea
F_3(\alpha,\alpha^\prime,\beta,\beta^\prime,\gamma,x,y)= \sum
{(\alpha)_m,(\alpha^\prime)_n (\beta)_m(\beta^\prime)_n \over
  (\gamma)_{m+n} m!n!} x^m y^n,
\ena
and $(\alpha)_n$ denotes $\Gamma(\alpha+n)/\Gamma(\alpha)$. We analytically
continue this function to the whole $x$-plane. Since the true value of
$x$ stays on the cut, we consider the value to be
\bea
            G_l (x) = \half(G_l(x+i0)+G_l(x-i0)),
\ena
which guarantee the validity of this integral around $x=0$.

We are now going to obtain the Bogoliubov coefficient. We first perform
some analytic continuations to rewrite the $F_3$ in (4.9) into ${}_2F_1$.
To begin with, we analytically continue the function
$F_3(\alpha,\alpha^\prime, \beta,\beta^\prime,\gamma,x,y)$ into the
function $F_2$ as\cite{F3formula}
\bea
&\quad& F_3(\alpha,\alpha^\prime,\beta,\beta^\prime,\gamma,
           {1 \over x^2},{1\over 1-x^2})\nonumber\\
 &=& \sum {\Gamma(\gamma)\Gamma(\rho-\lambda)\Gamma(\sigma-\mu) \over
           \Gamma(\rho)\Gamma(\sigma)\Gamma(\gamma-\Lambda-\mu)}
          (-x^{-2})^{-\lambda}(x^2-1)^{-\mu} \nonumber\\
&{}&\times F_2(\lambda+\mu+1-\gamma,\lambda,\mu,
               \lambda-\rho+1,\mu-\sigma+1;x^2,1-x^2),
\ena
where the sum consists of four terms in which $\lambda,\rho$ take
$\alpha,\beta; \beta,\alpha$ and $\mu,\rho$ is
$\alpha^\prime,\beta^\prime;\beta^\prime,\alpha^\prime$, and
$F_2(\alpha,\beta,\beta^\prime,\gamma,\gamma^\prime,x,y)$ is defined
by\cite{HTFFIIIDEF}
\bea
F_2(\alpha,\beta,\beta^\prime,\gamma,\gamma^\prime,x,y) = \sum
{(\alpha)_{m+n}(\beta)_m(\beta^\prime)_n \over
  (\gamma)_m(\gamma^\prime)_n m!n!} x^m y^n.
\ena
We then transform this function by using the formula\cite{HTFFIIAnalytic};
\bea
F_2(\alpha,\beta,\beta^\prime,\gamma,\gamma^\prime;x,y)
 =(1-x)^{-\alpha}
  F_2(\alpha,\gamma-\beta,\beta^\prime,\gamma,\gamma^\prime;
      {x \over  x-1},{y \over 1-x}),
\ena
which implies
\bea
&{}&F_3(\alpha,\alpha^\prime,\beta,\beta^\prime,\gamma;{1\over x^2},{1 \over
  1-x^2}) \nonumber\\
&=&  \sum e^{-\pi i \lambda} \cos\pi\mu
{\Gamma(\gamma)\Gamma(\rho-\lambda)\Gamma(\sigma-\mu) \over
   \Gamma(\rho)
   \Gamma(\sigma)\Gamma(\gamma-\Lambda-\mu)}x^{2
   \lambda}(1-x^2)^{\gamma-\lambda-1} \nonumber\\
&{}&\times F_2(\lambda+\mu+1,\lambda,\mu,\lambda-\rho+1;{x^2 \over
  x^2-1},1) ,
\ena
By the definition of $F_2$, we find
\bea
&{}&F_2(\alpha,\beta,\beta^\prime,\gamma,\gamma^\prime;x,1)
\nonumber\\
&{}& \qquad =
{\Gamma(\gamma^\prime)\Gamma(\gamma^\prime-\alpha -\beta^\prime) \over
  \Gamma(\gamma^\prime-\beta^\prime)\Gamma(\gamma^\prime-\alpha)}
{}_3F_2(\alpha,\beta,\alpha-\gamma^\prime+1;
        \gamma,\alpha+\beta^\prime-\gamma^\prime+1; x),
\ena
where ${}_3F_2(\alpha,\beta,\gamma;\delta,\lambda;x)$ is defined by
\bea
{}_3F_2(\alpha,\beta,\gamma;\delta,\lambda;x) = \sum_m
{(\alpha)_m(\beta)_m(\gamma)_m \over (\delta)_m(\lambda)_m m!}x^m.
\ena
Using this formula, we get
\bea
&{}&F_3(\alpha,\alpha^\prime,\beta,\beta^\prime,\gamma;
       {1 \over x^2},{1 \over  1-x^2})\nonumber\\
&=&\sum e^{-\pi i \lambda} \cos(\mu\pi)
       {\Gamma(\gamma)\Gamma(\rho-\lambda)\Gamma(\sigma-\mu)
        \Gamma(\mu-\sigma+1)\Gamma(\gamma-\lambda-\mu-\sigma) \over
        \Gamma(\rho)\Gamma(\sigma)\Gamma(\gamma-\Lambda-\mu)
        \Gamma(1-\sigma)\Gamma(\gamma-\lambda - \sigma)}\nonumber\\
&\times& x^{2\lambda}(1-x^2)^{\gamma-\lambda-1}
  {}_3F_2(\lambda+\mu-\gamma+1,\lambda+\sigma-\gamma+1,1-\rho;\nonumber\\
&{}&\hskip 120pt\lambda-\rho+1,\lambda+\mu+\sigma-\gamma+1;{x^2\over x^2-1}).
\ena
In our case, the parameters $(\alpha,\alpha^\prime,\beta,\beta^\prime,
\gamma)$ satisfy the following relation,
\bea
         \alpha+\alpha^\prime+\beta+\beta^\prime = \gamma,
\ena
so that we have
\bea
        1-\rho = \lambda + \mu + \sigma -\gamma +1.
\ena
Using this relation, we can find that the function reduces to
the sum of hypergeometric functions;
\bea
&{}&F_3(\alpha,\alpha^\prime,\beta,\beta^\prime,\gamma;
       {1 \over x^2},{1 \over  1-x^2})\nonumber\\
&=& e^{-\pi i \lambda}{ cos\mu\pi \sin\sigma\pi \over
     \sin(\sigma-\mu)\pi}{\Gamma(\gamma)\Gamma(\rho-\lambda) \over
                          \Gamma(\rho + \sigma)\Gamma(\rho+\mu)}
     x^{2\lambda}(1-x^2)^{\gamma-\lambda-1} \nonumber\\
&{}&\times F(1-\rho-\sigma, 1- \rho-\mu;\lambda-\rho+1;
            {x^2 \over  x^2-1}).
\ena
Making use of the relation;
\bea
F(a,b;c;z) = (1-z)^{c-a-b}F(c-a,c-b;c;z),
\ena
we finally obtain the formula
\bea
&{}&F_3(\alpha,\alpha^\prime,\beta,\beta^\prime,\gamma;
       {1 \over x^2},{1 \over 1-x^2})\nonumber\\
&=& e^{-\pi i \alpha}{\Gamma(\gamma)\Gamma(\beta-\alpha) \over
      \Gamma(\beta+\alpha^\prime)\Gamma(\beta+\beta^\prime)}
        \Bigl({x^{2}\over 1-x^2}\Bigr)^{\lambda}
   F(\alpha+\alpha^\prime, \alpha+\beta^\prime;\alpha-\beta+1;
       {x^2 \over  x^2-1})\nonumber\\
&{}& + (\alpha \leftrightarrow \beta),
\qquad(\hbox{for}\quad\alpha+\beta+\alpha^\prime+\beta^\prime=\gamma).
\ena

Next, we apply this formula to $(4.9)$ and obtain
\bea
G_l &=& e^{-{\pi \omega \over 2}}(1-x^2)^{-\half l}\Bigl[{
\Gamma(c+\half)\Gamma(\half) \over \Gamma({c \over 2} +\half
  +{\nu \over 2})\Gamma({c \over 2} +\half
  -{\nu \over 2})}F({c \over 2} +{\nu \over 2},{c \over 2}-{\nu \over
  2};\half;{x^2 \over x^2-1}) \nonumber\\
&+& {\Gamma(c+\half)\Gamma(-\half) \over \Gamma({c \over 2} + {\nu
    \over 2}) \Gamma({c \over 2} - {\nu
    \over 2})}e^{-{\pi \over 2}i}\Bigl( { x^2 \over 1 -x^2}
\Bigr)^\half  F({c \over 2} +{\nu \over 2} +\half ,{c \over 2}-{\nu \over
  2} +\half ;{3\over2};{x^2 \over x^2-1})\Bigr],
\ena
where we can observe the well-known factor $e^{-{\pi\omega \over
2}}$\cite{Haw} causing the Hawking radiation\cite{Lap}. Furthermore, we
can transform this function to a rather simple form by using the following
formulae;
\bea
&{}& {2 \Gamma(\half)\Gamma(a+b+\half) \over
  \Gamma(a+\half)\Gamma(b+\half)} F(a,b;\half;z) \nonumber\\
&=& F(2a,2b;a+b+\half;\half(1+z^\half)) +
F(2a,2b;a+b+\half;\half(1-z^\half)),\nonumber\\
&{}& {2\Gamma(-\half)\Gamma(a+b+\half) \over
\Gamma(a-\half)\Gamma(b-\half)} z^\half F(a,b;{3\over2};z) \nonumber\\
&=& F(2a-1,2b-1;a+b-\half; \half(1-z^\half)) \nonumber\\
&{}& - F(2a-1,2b-1;a+b-\half;\half(1+ z^\half)),
\ena
as
\bea
G_l = e^{-{\pi\omega \over2}}\sin^{-l}\eta_2
F(l+{3\over2}+\nu,l+{3\over2}-\nu,l+2;{ i e^{-i\eta_2}\over 2\sin\eta_2}),
\ena
and again using the formula $(4.22)$ we get
\bea
G_l &=& e^{-{\pi \omega \over 2}}(\sin\eta_2)(2i)^{l+1}e^{-i(l+1)\eta_2}
    F(\nu+ \half, -\nu + \half;l+2; { ie^{-i\eta_2} \over 2 \sin\eta_2})
\nonumber\\
&=& e^{-{\pi \omega \over 2}}\pi^{-\half}2^{l+{3\over2}}i^{l+1}
    e^{{i\over4}\pi -i\pi\nu}{\Gamma(l+2) \over
    \Gamma(l+\nu+{3\over2})}(\sin\eta_2)^{3\over2}
Q^\nu_{l+\half}(\cos\eta_2 + i0),
\ena
Note that the term which depends on $\omega$ is just the factor
$e^{-{\pi \omega \over 2}}$.

Quite similarly, we can find that the next term of the Bogoliubov
coefficient $G_{l+1}$ can be expressed as
\bea
G^{\pm}_{l+1} &=& \mp {i\omega \over 2(l+1)}
x^{i\omega-1}(1-x^2)^{-\half(l+i\omega)} \nonumber\\
&{}& \times F_3(-{i\omega \over 2} +
  \half,a,-{i\omega \over 2} +1,b;c+{3\over2},{1 \over x^2},{1 \over
    1-x^2}).
\ena
By using the formula $(4.23)$ again on this function, we obtain
\bea
G^\pm_{l+1}= {i\omega \over l+1}e^{-{\pi \omega \over
             2}}\pi^{-\half}2^{l+{3\over2}}
             i^{l+1}e^{{i\over4}\pi-i\pi\nu}{\Gamma(l+3) \over
             \Gamma(l+\nu+{5\over2})}(\sin\eta_2)^{3\over2}
             Q^\nu_{l+{3\over2}}(\cos\eta_2 + i0).
\ena
We confirm the relation $(4.5)$ comparing the coefficients of the mode
functions on the RHS of (4.27) and (4.29).
\setcounter{equation}{0}
\section{Discussions}

We have obtained Bogoliubov coefficient of the scalar wave functions
from the global coordinates to the static coordinates. We considered two
type of global coordinates, which we denoted type-I and type-II global
coordinates.

Technically, we have used rather different formula for each case. For the
type-I case, the Bogoliubov transformation corresponds to the
Weber-Schafheitlin type integral, whereas we have used analytic continuation
of the hypergeometric functions in two variables and proved a useful
formula $(4.23)$. The main difference of the correspondence of the
Bogoliubov transformation between the type-I and the type-II coordinates
stems from the use of the analytic continuations. In the type-I coordinates,
we did not use any analytic continuation which is required for type-II
coordinates. Unfortunately, we have not clarified the meaning of the analytic
continuation. We do not know  how to extract the famous factor
$e^{-{\pi\omega \over 2}}$ from the type-I case. To make sure the
difference between these two coordinates, let us argue the relation of the
mode functions in the type-I and type-II coordinates where the  relation of
the coordinates is given by
\bea
- \eta_1 = { \sin\eta_2 \over \cos\chi - \cos\eta_2},\qquad r_1 =
{\sin\chi \over \cos\chi-\cos\eta_2},
\ena
for $\cos\chi>\cos\eta_2$.

Since there seems to be any direct formula in this case, we start with
the following integral formula\cite{TITI};
\bea
&{}&\int_0^\infty x^{\rho+\nu-\mu+1}J_\mu (x)J_\nu(\beta x)K_\rho(\gamma
x) dx \nonumber\\
&=& 2^{\rho+\nu-\mu-1}{\Gamma(\rho+\nu+1)\Gamma(\rho+1)\Gamma(\nu+1)
  \over \Gamma(\mu+1)}(\cosh\sigma - \cos\theta) \nonumber\\
&{}&\times
P^{-\rho}_{\rho+\nu-\mu}(\cos\theta)P^{-\nu}_{\rho+\nu-\mu}(\cosh\sigma),
\ena
where
\bea
\beta = {\sinh\sigma \over \cosh\sigma - \cos\theta}, \qquad \gamma =
{\sin\theta \over \cosh\sigma - \cos\theta}.
\ena
This integral is also Weber-Schafheitlin type. We could not find any
proof of this formula in the literature. The proof may require some
factorization formula of $F_4$ which is one of the hypergeometric
function of two variables\cite{HTFFIIIDEF}.

We will make an analytic continuation by $ \sigma \rightarrow i
\sigma$ keeping a relation $\cos\theta < \cos\sigma$, intrinsic to
discontinuous integral. Then we have
\bea
&{}&\int_0^\infty x^{\rho+\nu-\mu+1}J_\mu (x)I_\nu(\beta x)K_\rho(\gamma
x) dx \nonumber\\
&=& 2^{\rho+\nu-\mu-1}{\Gamma(\rho+\nu+1)\Gamma(\rho+1)\Gamma(\nu+1)
  \over \Gamma(\mu+1)}(\cos\sigma - \cos\theta) \nonumber\\
&{}&\times
P^{-\rho}_{\rho+\nu-\mu}(\cos\theta)P^{-\nu}_{\rho+\nu-\mu}(\cos\sigma),
\ena
where
\bea
\beta = {\sin\sigma \over \cos\sigma - \cos\theta}, \qquad \gamma =
{\sin\theta \over \cos\sigma - \cos\theta}.
\ena
In the integral, we will use another integral formula\cite{TITII}
\bea
I_\nu(\beta x) K_\rho(\gamma x)
&=& \int_0^\infty dt {x^{-\lambda} t^{\lambda+1} \over t^2 + x^2}
\lbrace cos[\half(\lambda-\rho+\nu)\pi]J_\rho(\gamma t) \nonumber\\
&{}& +
\sin[\half(\lambda-\rho + \nu)\pi]Y_\rho(\gamma t) \rbrace J_\nu(\beta
t),
\ena
and setting $\lambda = \rho +\nu - 2\mu$, performing integration over
$x$ by using a formula\cite{TITIII},
we have the following integral formula
\bea
&{}& \int_0^\infty dk k^{N+\half} K_{-\nu-N+l}(k) \lbrace
\cos[(N+\nu+\half)\pi] \rbrace J_{-\nu}(-k\eta_1) \nonumber\\
&{}& \quad + \sin[(N+\nu+\half)\pi] Y_{-\nu}(-k\eta_1) \rbrace
(-k\eta_1)^{3\over2} J_{l+\half}(kr_1)(kr_1)^{-\half} \nonumber\\
&=& 2^{N-l-1}(-1)^{-\half l - {1 \over
    4}}{\Gamma(l-\nu+{3\over2})\Gamma(-\nu+1)
  \Gamma(2l+2)\Gamma(N-l+1) \over \Gamma(-\nu-N+l+1)
  \Gamma(N+l+2)} \nonumber\\
&{}& \times (\sin\eta_2)^{3 \over2} P^{\nu}_{N+\half}(\cos\eta_2) (\sin\chi)^l
C^{l+1}_{N-l}(\cos\chi).
\ena

Unfortunately, from this kind of Bogoliubov transformation, we can not
fix the Bogoliubov transformation uniquely because there are many
vanishing identities in the integral formula. As a matter of fact,
we can change the ratio of $J_\rho$ and $Y_\rho$ by  the choice of the
parameter $\lambda$ in $(5.6)$. Therefore we are required to consider the
inverse transformation of $(5.7)$ in order to fix the coefficient, which we
fail to perform. Anyway, the formula $(5.7)$ shows an existence of the
Bogoliubov transformation between the mode functions of global coordinates,
which is expected because we know the Green functions for both global
coordinates are identical.

In this paper, we have obtained the Bogoliubov coefficients from global
coordinates to the static coordinates. Although we have shown that the
Bogoliubov transformations are related to some mathematical formula
including those of hypergeometric functions in two variables, the true
physical interpretation of the analytic continuation is not clear,
which deserves further investigation.
%


\newpage
\begin{thebibliography}{99}
%
\bibitem{Haw}
S.W.Hawking, Commun.Math.Phys. 43 (1975) 199.
%
\bibitem{Nac}
O.Nachtmann, Commn.Math.Phys. 6 (1967) 1.
%
\bibitem{CT}
N.A. Chernikow and E.A. Tagirov,
Ann. Inst. Henri Poincar$\acute e$, 9A (1968) 109.
%
\bibitem{RP}
D. Rohiya and N. Panchapakesan, J. Phys. A: Gen.Phys., 11 (1978) 1963.
%
\bibitem{Tag}
E.A. Tagirov, Ann. Phys. 76 (1973) 561.
%
\bibitem{GH}
G.W. Gibbons and S.W. Hawining, Phys. Rev. D15 (1977) 2738.
%
\bibitem{Unr}
W.G.Unruh, Phys.Rev.D14(1976) 870.
%
\bibitem{Lap}
A.S. Lapedes, J. Math. Phys. 19 (1987) 2289.
%
\bibitem{HTFI}
`` Higher Transcendental Functions,'' eds. A.Erd{\'e}lyi et al.,
McGraw-Hill, New York (1953) Vol.II, p 51 (29).
%
\bibitem{HTFII}
``Higher Transcendental Functions,'' eds. A.Erd{\'e}lyi et al.,
McGraw-Hill, New York (1953) Vol.I, p 101.
%
\bibitem{HTFFIIIINT}
`` Higher Transcendental Functions,''eds. A.Erd{\'e}lyi et al.,
McGraw-Hill, New York (1953) Vol.I, p.231,(7).
%
\bibitem{F3formula}
P.Appell and M.J.Kamp$\acute e$ de F$\acute e$riet,
``Fonctions hyperg{\'e}om{\'e}triques et hypersph{\'e}riques,'' Polymones
d`Hermite, Gauthier-Villars (1926);\\
A. Erd{\'e}lyi, Proc. Roy. Soc. Edinburgh Sect. A62 (1948) pp.378-385;\\
``Higher Transcendental Functions,''  eds. A.Erd{\'e}lyi et al.,
McGraw-Hill, New York (1953) Vol.I, p.241,(11).
%
\bibitem{HTFFIIIDEF}
`` Higher Transcendental Functions,'' eds. A.Erd{\'e}lyi et al.,
McGraw-Hill, New York (1953) Vol.I. p 224.
%
\bibitem{HTFFIIAnalytic}
``Higher Transcendental Functions,'' eds. A.Erd{\'e}lyi et al.,
McGraw-Hill, New York (1953) Vol.I. p.240,(6).
%
\bibitem{TITI}
``Table of Integral Transforms,'' eds. A.Erd{\'e}lyi et al.,
McGraw-Hill, New York (1954) Vol.II, p65,(14).
%
\bibitem{TITII}
``Table of Integral Transforms,'' eds. A.Erd{\'e}lyi et al.,
McGraw-Hill, New York (1954) Vol.II, p55,(47).
%
\bibitem{TITIII}
``Table of Integral Transforms,'' eds. A.Erd{\'e}lyi et al.,
McGraw-Hill, New York (1954)Vol.II, p23,(12).
%
\end{thebibliography}
\end{document}